\title{Exact Enumeration and Sampling of Matrices with Specified Margins}

\author{
Jeffrey W. Miller
\thanks{
Division of Applied Mathematics, Brown University, Providence, RI 02912.
Email: jeffrey\_miller at brown.edu.
Research supported by a NDSEG fellowship and in part by NSF award DMS-1007593.} 
\and  
Matthew T. Harrison
\thanks{
Division of Applied Mathematics, Brown University, Providence, RI 02912.
Research supported by NSF award DMS-1007593.}
}

\date{}

\documentclass[12pt,letterpaper]{article}

\usepackage[]{graphicx}
\usepackage{enumitem}
\usepackage{times}      
\usepackage{amssymb}
\usepackage{amsmath}
\usepackage[usenames]{color}
\usepackage{verbatim}

\usepackage{ifthen}
\newboolean{details}
\setboolean{details}{false}

\newcommand{\qed}{\nobreak \ifvmode \relax \else
      \ifdim\lastskip<1.5em \hskip-\lastskip
      \hskip1.5em plus0em minus0.5em \fi \nobreak
      \vrule height0.75em width0.5em depth0.25em\fi}
\newtheorem{thm}{Theorem}[section]
\newtheorem{lem}[thm]{Lemma}

\newtheorem{cor}[thm]{Corollary}
\newtheorem{detail}[thm]{Detail}
\newtheorem{alg}[thm]{Algorithm}
\newenvironment{pf}[1][Proof\,]{\begin{trivlist}
\item[\hskip \labelsep {\bfseries #1}]}{\qed\end{trivlist}}

\newenvironment{rk}[1][Remark\,]{\begin{trivlist}
\item[\hskip \labelsep {\bfseries #1}]}{\end{trivlist}}

\newcommand{\ds}{\displaystyle}

\renewcommand{\u}{\cup}

\newcommand{\hs}{\hspace{1mm}}
\newcommand{\nl}{\vspace{5mm}}
\newcommand{\nnl}{\nl\nl}

\newcommand{\x}{\times}

\renewcommand{\b}{\beta}

\renewcommand{\d}{\Delta}

\renewcommand{\t}{\tau}

\newcommand{\for}{\mbox{\, for \,}}

\renewcommand{\And}{\mbox{\, and \,}}
\newcommand{\sbs}{\subset}
\newcommand{\sps}{\supset}

\newcommand{\inv}{^{-1}}

\renewcommand{\O}{\textup{O}}

\newcommand{\todo}[1]{\emph{{(#1)}}}

\newcommand{\enum}[1]{\begin{enumerate}[topsep=-0pt,partopsep=0pt,parsep=0pt,itemsep=0pt] #1\end{enumerate}}
\newcommand{\bull}[1]{\begin{itemize}[topsep=-0pt,partopsep=0pt,parsep=0pt,itemsep=0pt] #1\end{itemize}}

\newcommand{\mat}[1]{\begin{matrix}#1\end{matrix}}

\newcommand{\branch}[4]{
\left\{
	\begin{array}{ll}
		#1  & \mbox{if } #2 \\
		#3 & \mbox{if } #4
	\end{array}
\right.
}

\newcommand{\eq}[1]{\stackrel{\textup{(#1)}}{=}}
\newcommand{\leql}[1]{\stackrel{\textup{(#1)}}{\leq}}

\newcommand{\D}{\mathcal{D}}

\DeclareSymbolFont{AMSb}{U}{msb}{m}{n}
\DeclareMathSymbol{\N}{\mathbin}{AMSb}{"4E}
\DeclareMathSymbol{\Z}{\mathbin}{AMSb}{"5A}
\DeclareMathSymbol{\R}{\mathbin}{AMSb}{"52}
\DeclareMathSymbol{\Q}{\mathbin}{AMSb}{"51}
\DeclareMathSymbol{\C}{\mathbin}{AMSb}{"43}
\DeclareMathSymbol{\G}{\mathbin}{AMSb}{"47}
\DeclareMathSymbol{\B}{\mathbin}{AMSb}{"42}
\DeclareMathSymbol{\Set}{\mathbin}{AMSb}{"53}
\DeclareMathSymbol{\BA}{\mathbin}{AMSb}{"41}
\DeclareMathSymbol{\BB}{\mathbin}{AMSb}{"42}

\renewcommand{\b}{\bar }
\newcommand{\p}{{\bf p}}
\newcommand{\q}{{\bf q}}
\renewcommand{\r}{{\bf r}}
\newcommand{\s}{{\bf s}}
\newcommand{\ttt}{{\bf t}}
\renewcommand{\u}{{\bf u}}
\renewcommand{\v}{{\bf v}}

\renewcommand{\r}{{\bf r}}

\newcommand{\xx}{{\bf x}}
\newcommand{\X}{{\bf X}}
\newcommand{\0}{{\bf 0}}
\newcommand{\1}{{\bf 1}}
\newcommand{\2}{{\bf 2}}

\newcommand{\bq}{\bar{\bf q}}
\newcommand{\br}{\bar{\bf r}}

\newcommand{\bv}{{\bar{\bf v}}}

\newcommand{\bN}{\bar N}
\newcommand{\bM}{\bar M}
\newcommand{\reduce}{\backslash}

\newcommand{\Nn}{\N^n}
\newcommand{\Nm}{\N^m}

\newcommand{\Bn}{\B^n}
\newcommand{\del}{\partial}
\renewcommand{\H}{H_n(r)}
\newcommand{\Hs}{H^*_n(r)}

\begin{document} 
\maketitle

\begin{abstract}
We describe a dynamic programming algorithm for exact counting and exact uniform sampling of matrices with specified row and column sums. The algorithm runs in polynomial time when the column sums are bounded. Binary or non-negative integer matrices are handled. The method is distinguished by applicability to non-regular margins, tractability on large matrices, and the capacity for exact sampling.

\nl {\footnotesize Keywords: bipartite graphs, specified degrees, exact counting, exact sampling, tables}

\end{abstract}


\section{Introduction}

Let $N(\p,\q)$ be the number of $m\x n$ binary matrices
with margins (row and column sums) $\p=(p_1,\dotsc,p_m)\in\N^m$, $\q=(q_1,\dotsc,q_n)\in\N^n$
respectively, and let $M(\p,\q)$ be the corresponding number of $\N$-valued matrices.
In this paper we develop a technique for efficiently finding $N(\p,\q)$ and $M(\p,\q)$.
Uniform sampling from these sets of matrices is an important problem in statistics
\cite{chen05},
and the method given here permits efficient exact uniform sampling
once the underlying enumeration problem has been solved.

Since a bipartite graph with degree sequences $\p=(p_1,\dotsc,p_m)\in\N^m$, $\q=(q_1,\dotsc,q_n)\in\N^n$
(and $m,n$ vertices in each part respectively) can be viewed as a $m\times n$ matrix with row and column sums $(\p,\q)$, our technique applies equally well to counting and uniformly sampling such bipartite graphs. Under this correspondence, simple graphs correspond to binary matrices, and multigraphs correspond to $\N$-valued matrices.

The distinguishing characteristic of the method is its tractability on matrices of non-trivial size.
In general, computing $M(\p,\q)$ is \#P-complete \cite{dyer97},
and perhaps $N(\p,\q)$ is as well.
However, if we assume a bound on the column sums 
then our algorithm computes both numbers in polynomial time.
After enumeration, uniform samples may be drawn in polynomial expected time for bounded column sums.
To our knowledge, all previous algorithms {\it for the non-regular case}
require super-polynomial time (in the worst case) to compute these numbers, even for bounded column sums.
(We assume a description length of at least $m+n$ 
and no more than $m\log a+n\log b$, where $a=\max p_i$, $b=\max q_i$.)
In general (without assuming a bound on the column sums),
our algorithm computes $N(\p,\q)$ or $M(\p,\q)$ in $\O(m(ab+c)(a+b)^{b-1}(b+c)^{b-1}(\log c)^3)$ time
for $m\x n$ matrices,
where $a=\max p_i$, $b=\max q_i$, and $c=\sum p_i=\sum q_i$.
After enumeration, uniform samples may be drawn in $\O(mc\log c)$ expected time.

In complement to most approaches to computing $M(\p,\q)$, which are efficient for small matrices with large margins, our algorithm is efficient for large matrices with small margins.
For instance, in Section \ref{sec:app} we count the $100\times 100$ matrices with margins
$(70,30,20,10,5^{(6)},4^{(10)},3^{(20)},2^{(60)})$, $(4^{(80)},3^{(20)})$ 
(where $x^{(n)}$ denotes $x$ repeated $n$ times).

To illustrate the problem at hand, consider a trivial example: if
$\p=(2,2,1,1)$, $\q=(3,2,1)$, then $N(\p,\q)=8$ and $M(\p,\q)=24$.
The $8$ binary matrices are below.
$${\footnotesize
      \mat{1&1&0\\1&1&0\\1&0&0\\0&0&1}
\hspace{6mm}\mat{1&1&0\\1&1&0\\0&0&1\\1&0&0}
\hspace{6mm}\mat{1&1&0\\1&0&1\\1&0&0\\0&1&0}
\hspace{6mm}\mat{1&1&0\\1&0&1\\0&1&0\\1&0&0}
\hspace{6mm}\mat{1&1&0\\0&1&1\\1&0&0\\1&0&0}
\hspace{6mm}\mat{1&0&1\\1&1&0\\1&0&0\\0&1&0}
\hspace{6mm}\mat{1&0&1\\1&1&0\\0&1&0\\1&0&0}
\hspace{6mm}\mat{0&1&1\\1&1&0\\1&0&0\\1&0&0}
}
$$

The paper will proceed as follows:
\bull{
\item[\S2] Main results
\item[\S3] Brief review
\item[\S4] Applications
\item[\S5] Proof of recursions
\item[\S6] Proof of bounds on computation time.
}

\section{Main results: Recursions, Bounds, Algorithms}
\label{sec:rec}

Introducing the following notation will be useful.
Taking $\N:=\{0,1,2,\dotsc\}$, we consider $\Nn$ to be the subset of
$\N^\infty :=\{(r_1,r_2,\dotsc): r_i\in\N \for i = 1,2,\dotsc\}$ 
such that all but the first $n$ components are zero.
Let $L:\N^\infty\to\N^\infty$ denote the left-shift map: $L\r=(r_2,\dotsc,r_n,0,0,\dotsc)$.
Given $\r,\s \in \Nn$, let
$\r \reduce \s := \r-\s+L\s$, (which may be read as ``$\r$ {\it reduce} $\s$''), let
$$
{\r\choose\s} := {r_1\choose s_1} \cdots {r_n\choose s_n},
$$
and let $\br$ denote the vector of counts,
$\br := (\bar r_1,\bar r_2,\dotsc)$ where $\bar r_i:=\#\{j:r_j=i\}$.
We write $\r\le\s$ if $r_i\le s_i$ for all $i$.
Given $n\in\N$, let $C_n(k):=\{\r\in\Nn : \sum_i r_i=k\}$ be the $n$-part compositions (including zero) of $k$,
and given $\s\in\N^n$, let $C^\s(k):=\{\r\in C_n(k) : \r\le\s \}$.
For $(\p,\q)\in\Nm\x\Nn$, define the numbers
$$
N(\p,\q):=\#\{\X\in\{0,1\}^{m\times n}:\sum_j x_{ij}=p_i,\sum_i x_{ij}=q_j, \for 1\leq i\leq m, 1\leq j\leq n\},
$$
$$
M(\p,\q):=\#\{\X\in\N^{m\times n}:\sum_j x_{ij}=p_i,\sum_i x_{ij}=q_j, \for 1\leq i\leq m, 1\leq j\leq n\}.
$$
Since $N(\p,\q)$ and $M(\p,\q)$ are fixed under permutations
of the row sums $\p$ and column sums $\q$, and since zero margins do not affect the number of matrices and can effectively be ignored, then we may define
$\bN(\p,\bq):=N(\p,\q)$ and $\bM(\p,\bq):=M(\p,\q)$ without ambiguity.
We can now state our main results.

\begin{thm}[Recursions]
\label{thm:rec}
The number of matrices with margins $(\p,\q)\in\Nm\x\Nn$ is given by
\enum{
\item $\ds\bN(\p,\r)=\sum_{\s\in C^\r(p_1)} {\r\choose\s} \bN(L\p,\r\reduce\s)$
\hspace{6mm} for binary matrices, and
\item $\ds\bM(\p,\r)=\sum_{\s\in C^{\r+L\s}(p_1)} {\r+L\s \choose\s} \bM(L\p,\r\reduce\s)$
\hspace{6mm} for $\N$-valued matrices,
}
where $\r=\bq$, and in \textup{(2)}, we sum over all $\s$ such that $\s\in C^{\r+L\s}(p_1)$.
\end{thm}

Proofs will be given in Section \ref{sec:pf}.
The Gale-Ryser conditions \cite{gale57,ryser57} simplify computation of the sum in (1) 
by providing a necessary and sufficient condition for there to exist a binary matrix with margins $(\p,\q)$:
if $q_i':=\#\{j : q_j\geq i\}$ and $p_1\geq\cdots\geq p_m$, then $N(\p,\q)\neq 0$ 
if and only if $\sum_{i=1}^j p_i \leq \sum_{i=1}^j q_i' \mbox{ for all } j<m
\And \sum_{i=1}^m p_i = \sum_{i=1}^m q_i'.$  
This is easily translated into a similar condition in terms of $(\p,\bq)$ and $\bN(\p,\bq)$.
The following recursive procedure can be used to compute either $N(\p,\q)$ or $M(\p,\q)$.

\begin{alg}[Enumeration] \label{alg:c} \hs\\
Input: $(\p,\bq)$, where $(\p,\q)\in\Nm\x\Nn$ are row and column sums such that $\sum_i p_i=\sum_i q_i$.\\
Output: $N(\p,\q)$ (or $M(\p,\q)$), the number of binary (or $\N$-valued) matrices.\\
Storage: Lookup table of cached results, initialized with $\bN(\0,\0)=1$ (or $\bM(\0,\0)=1$).
\enum{
\item If $\bN(\p,\bq)$ is in the lookup table, return the result.
\item In the binary case, if Gale-Ryser gives $\bN(\p,\bq)=0$, cache the result and return 0.
\item Evaluate the sum in Theorem \ref{thm:rec}, recursing to step \textup{(1)} for each term.
\item Cache the result and return it.
}
\end{alg}

Let $T(\p,\q)$ be the time (number of machine operations) required by Algorithm \ref{alg:c} to compute
$N(\p,\q)$ or $M(\p,\q)$, after performing an $\O(n^3)$ preprocessing step to compute all needed binomial coefficients.
(It turns out that computing $M(\p,\q)$ always takes longer, but the bounds we prove apply to both $N(\p,\q)$ and $M(\p,\q)$.)
We give a series of bounds on $T(\p,\q)$ ranging from tighter but more complicated,
to more crude but simpler. The bounds will absorb the $\O(n^3)$ pre-computation except for the trivial case when the maximum column sum is $1$.

\begin{thm}[Bounds] Suppose $(\p,\q)\in\Nm\x\Nn$, $a=\max p_i$, $b=\max q_i$, and $c=\sum p_i =\sum q_i$. Then
\label{thm:eff}
\enum{
\item 
$\ds T(\p,\q) \leq \O((ab+c)(\log c)^3\sum_{i=1}^m {p_i+b-1 \choose b-1}{p_i+\cdots+p_m + b-1 \choose b-1}),$
\item $\ds T(\p,\q) \leq \O(m(ab+c)(a+b)^{b-1}(b+c)^{b-1}(\log c)^3),$
\item $\ds T(\p,\q) \leq \O(mn^{2b-1}(\log n)^3)$ for bounded $b$,
\item $\ds T(\p,\q) \leq \O(mn^{b}(\log n)^3)$ for bounded $a,b$.
}
\end{thm}
\begin{rk}
Since we may swap the row sums with the column sums without changing the number of matrices,
we could use Algorithm \ref{alg:c} on $(\q,\bar\p)$ to compute $N(\p,\q)$ or $M(\p,\q)$ using $T(\q,\p)$ operations, which, for example, is $\O(nm^{a}(\log m)^3)$ for bounded $a,b$.
$T(\p,\q)$ also depends on the ordering of the row sums $p_1,\dotsc,p_m$ as suggested by Theorem \ref{thm:eff}(1), and we find that putting them in decreasing order $p_1\geq\cdots\geq p_m$ tends to work well.
Algorithm \ref{alg:c} is typically made significantly more efficient
by using the Gale-Ryser conditions, and this is not accounted for in these bounds.
Although we observe empirically that this reduces computation tremendously, 
we do not have a proof of this.  
\end{rk}

Algorithm \ref{alg:c} traverses a directed acyclic graph
in which each node represents a distinct set of input arguments to the algorithm, such as $(\p,\bq)$.
Node $(\u,\bv)$ is the child of node $(\p,\bq)$ if the algorithm is called (recursively) with 
arguments $(\u,\bv)$ while executing a call with arguments $(\p,\bq)$.
If the initial input arguments are $(\p,\bq)$, then all nodes are descendents of node $(\p,\bq)$.
Meanwhile, all nodes are ancestors of node $(\0,\0)$.
Note the correspondence between the children of a node $(\u,\bv)$ and the
compositions $\s\in C^{\bv}(u_1)$ in the binary case, and $\s\in C^{\bv+L\s}(u_1)$ in the $\N$-valued case,
under which $\s$ corresponds with the child $(L\u,\bv\reduce\s)$.
We also associate with each node its {\it count}:
the number of matrices with the corresponding margins.

As an additional benefit of caching the counts in a lookup table (as in Algorithm \ref{alg:c}),
once the enumeration is complete we obtain an efficient algorithm
for uniform sampling from the set of $(\p,\q)$ matrices (binary or $\N$-valued).
It is straightforward to prove that since the counts are exact,
the following algorithm yields a sample from the uniform distribution.

\begin{alg}[Sampling] \label{alg:s} \hs\\
Input:
\\ $\cdot$ Row and column sums $(\p,\q)\in\Nm\x\Nn$ such that $\sum_i p_i=\sum_i q_i$.
\\ $\cdot$ Lookup table of counts generated by Algorithm \ref{alg:c} on input $(\p,\bq)$. \\
Output: A binary (or $\N$-valued) matrix with margins $(\p,\q)$, drawn uniformly at random.
\enum{
\item Initialize $(\u,\v) \leftarrow (\p,\q)$.
\item If $(\u,\v)=(\0,\0)$, exit.
\item\label{choose} Choose a child $(L\u,\bv\reduce\s)$ of $(\u,\bv)$ with probability proportional to its count times the number of corresponding rows (that is, the rows $\r\in C^\v(u_1)$ such that
$\overline{\v-\r}=\bv\reduce\s$.)
\item\label{step} Choose a row uniformly among the corresponding rows.
\item $(\u,\v) \leftarrow (L\u,\v-\r)$.
\item Goto \textup{(2)}.
}
\end{alg} In step (\ref{choose}), there are ${\bv\choose\s}$ corresponding rows $\r$ in the binary case,
and ${\bv+L\s\choose\s}$ in the $\N$-valued case.
In step (\ref{step}), in the binary case of course we only choose among $\r\in\{0,1\}^n$.
In Section \ref{sec:comp} we prove that Algorithm \ref{alg:s}
takes $\O(mc\log c)$ expected time per sample, where $c=\sum_i p_i$.
\nnl

\section{Brief review}
\label{sec:rev}

We briefly cover the previous work on this problem.  
This review is not exhaustive, focusing instead on those results which are particularly significant
or closely related to the present work.
Let $\H$ and $\Hs$ denote $M(\p,\q)$ and $N(\p,\q)$, respectively, when $\p=\q=(r,\dotsc,r)\in\Nn$.
The predominant focus has been on the regular cases $\H$ and $\Hs$.

Work on counting these matrices goes back at least as far as MacMahon, who applied his expansive theory to find
the polynomial for $H_3(r)$ \cite{macmahon15} (Vol II, p.161), and developed the theory of Hammond
operators, which we will use below.
Redfield's theorem \cite{redfield27}, inspired by MacMahon, 
can be used to derive summations for some special cases,
such as $\H, \Hs$ for $r=2,3$, and 
in similar work Read \cite{read59,read60} used P\'{o}lya theory to derive these summations for $r=3$.
Two beautiful theoretical results must also be mentioned:
Stanley \cite{stan73} proved that for fixed $n$, $\H$ is a polynomial in $r$, and
Gessel \cite{gess87,gess90} showed that for fixed $r$, both $\H$ and $\Hs$ are P-recursive in $n$,
vastly generalizing the linear recursions for $H_n(2)$, $H^*_n(2)$ found by Anand, Dumir, and Gupta \cite{anand66}.

We turn next to algorithmic results more closely related to the present work.
McKay \cite{mckay83,canfield05} has demonstrated a coefficient extraction technique 
for computing $N(\p,\q)$ in the semi-regular case (in which $\p=(a,\dotsc,a)\in\Nm$ and $\q=(b,\dotsc,b)\in\Nn$). 
To our knowledge,
McKay's is the most efficient method known previously for $N(\p,\q)$.
By our analysis it requires at least $\Omega(mn^{b})$ time for bounded $a,b$,
while the method presented here is $\O(mn^{b}(\log n)^3)$ in this case.
Since this latter bound is quite crude, we expect that our method should have comparable or better performance, and indeed empirically we find that typically it is more efficient.
If only $b$ is bounded, McKay's algorithm is still $\Omega(mn^b)$, 
but the bound on our performance increases to $\O(mn^{2b-1}(\log n)^3)$,
so it is possible that McKay's algorithm will outperform ours in these cases.
Nonetheless, it is important to bear in mind that McKay's algorithm 
is efficient only in the semi-regular case (while our method permits non-regular margins).
If neither $a$ nor $b$ is bounded, McKay's method is exponential in $b$ (as is ours).

Regarding $M(\p,\q)$, one of the most efficient algorithms known to date is LattE (Lattice point Enumeration) \cite{deloera04}, which uses Barvinok's algorithm \cite{barvinok94} to count lattice points contained in convex polyhedra. It runs in polynomial time for any fixed dimension, and as a result it can compute $M(\p,\q)$ for astoundingly large margins, provided that $m$ and $n$ are small. However, since the computation time grows very quickly with the dimension, LattE is currently inapplicable when $m$ and $n$ are larger than $6$.
There are similar algorithms \cite{mount00,deloera03,beck03} that are efficient for small matrices. 

In addition, several other algorithms have been presented for finding 
$N(\p,\q)$ (such as \cite{johnsen87,wang88,wang98,perez02}) 
and $M(\p,\q)$ (see review \cite{diaconis95}) allowing non-regular margins,
however, it appears that all are exponential in the size of the matrix, even for bounded margins.
While in this work we are concerned solely with exact results,
we note that many useful approximations
for $N(\p,\q)$ and $M(\p,\q)$ (in the general case) have been found, 
as well as approximate sampling algorithms \cite{holmes96,chen05,greenhill06,canfield08,harrison09}.

\section{Applications}
\label{sec:app}

\subsection{Occurrence matrices from ecology}

The need to count and sample occurrence matrices
(binary matrices indicating observed pairings of elements of two sets)
arises in ecology.  A standard dataset of this type is ``Darwin's finch data'',
a $13\x 17$ matrix indicating which of 13 species of finches inhabit which of 17 of
the Gal\'{a}pagos Islands.  The margins of this matrix are
(14, 13, 14, 10, 12, 2, 10, 1, 10, 11, 6, 2, 17),
(4, 4, 11, 10, 10, 8, 9, 10, 8, 9, 3, 10, 4, 7, 9, 3, 3).
We count the number of such matrices to be 67,149,106,137,567,626 (in 1.5 seconds)
confirming \cite{chen05}.
Further, we sample exactly from the uniform distribution over this set 
at a rate of 0.001 seconds per sample.
(All computations were performed on a 64-bit 2.8 GHz machine with 6 GB of RAM.)
A similar dataset describes the distribution of 23 land birds
on the 15 southern islands in the Gulf of California \cite{chen05,cody83}:
this binary matrix has margins
(14, 14, 14, 12, 5, 13, 9, 11, 11, 11, 11, 11, 7, 8, 8, 7, 2, 4, 2, 3, 2, 2, 2),
(21, 19, 18, 19, 14, 15, 12, 15, 12, 12, 12, 5, 4, 4, 1),
for which we count 839,926,782,939,601,640 corresponding binary matrices.
Counting takes 1 second, and sampling is 0.002 seconds per sample.
One more example of this type: for bird species on the California Islands \cite{power72}
we find that there are 1,360,641,571,195,211,109,388 binary matrices with margins 
(1, 4, 3, 2, 1, 1, 1, 5, 1, 3, 1, 4, 4, 5, 1, 2, 1, 5, 4, 5, 3, 7, 1, 3, 2, 4, 1, 3, 2, 4, 6),
(2, 14, 24, 8, 2, 5, 20, 15) in 4 seconds; samples take 0.003 seconds each.
Larger matrices can be handled as well, provided the margins are small.
For example, we count 860585058801817078819959949756...000 (459 digits total, see Appendix)
$100\x 100$ matrices with margins $(70,30,20,10,5^{(6)},4^{(10)},3^{(20)},2^{(60)})$, $(4^{(80)},3^{(20)})$ 
(where $x^{(n)}$ denotes $x$ repeated $n$ times) in 46 minutes.  
We know of no previous algorithm capable of
efficiently and exactly counting and sampling from sets such as this.

\subsection{Ehrhart polynomials of the Birkhoff polytope}

Stanley \cite{stan73} proved a remarkable conjecture of Anand, Dumir, and Gupta \cite{anand66}:
given $n\in\N$, $H_n(r)$ is a polynomial in $r$ (where $H_n(r)=M(\p,\q)$ with $\p=\q=(r,r,\dotsc,r)\in\Nn$).  
Given $H_n(1),\dotsc,H_n({n-1 \choose 2})$, one can solve for the coefficients of $H_n(r)$ (as we describe below).
These polynomials have been computed for $n\leq 9$ by Beck and Pixton \cite{beck03}.
As an application of our method, we computed them for $n\leq 8$, and found that the computation time
is comparable to that of Beck and Pixton. 
For $n=4,\dotsc,8$, the numbers $H_n(r)$ for $r = 1,\dotsc,{n-1 \choose 2}$ are listed in the Appendix,
and the polynomial $H_4(r)$ is displayed here as an example. 
Our results confirm those of Beck and Pixton.

$$H_4(r)=1+(65/18) r+ (379/63) r^{2}+ (35117/5670) r^{3}+ (43/10) r^{4}$$
$$+(1109/540) r^{5}+ (2/3) r^{6}+ (19/135) r^{7}+ (11/630) r^{8}+ (11/11340) r^{9}.$$

The coefficients of $H_n(r)$ can be determined by the following method.
By Stanley's theorem \cite{stan73}, $H_n(r)$ is a polynomial in $r$
such that (a) $\deg H_n(r)=(n-1)^2$, (b) $H_n(-1)=\cdots=H_n(-n+1)=0$,
and (c) $H_n(-n-r)=(-1)^{(n-1)^2}H_n(r)$ for $r\in\N$.
For each $n=4,\dotsc,8$, we perform the following computation. 
Let $k ={n-1 \choose 2}$ and $d =(n-1)^2$.
Compute the numbers $H_n(r)$ for $r = 0,1,\dotsc,k$ 
using Algorithm \ref{alg:c},
and form the vector $\v :=(H_n(-n-k+1),\dotsc,H_n(k))^{\top}\in\Z^{d+1}$ 
using (b) and (c).
Form the matrix $A = \big((i-n-k)^{j-1}\big)_{i,j=1}^{d+1}\in\Z^{(d+1)\x(d+1)}$,
and compute $\u = A\inv\v$.  Then by (a),
$$H_n(r)=\sum_{j=0}^d u_{j+1} r^j.$$

\subsection{Contingency Tables}

As an example of counting contingency tables with non-regular margins, we count 620017488391049592297896956531...000 (483 digits total, see Appendix)
$100\x 100$ matrices with margins $(70,30,20,10,5^{(6)},4^{(10)},3^{(20)},2^{(60)})$, $(4^{(80)},3^{(20)})$ 
(where $x^{(n)}$ denotes $x$ repeated $n$ times) in 118 minutes.  
Again, we know of no previous algorithm capable of
efficiently and exactly counting and sampling from sets such as this.
(However, for small contingency tables with large margins, our algorithm is much less efficient than other methods such as LattE.)
Exact uniform sampling is possible for contingency tables as well, which occasionally finds use in statistics \cite{diaconis85}.

\section{Proof of recursions}
\label{sec:pf} 

We give two proofs of Theorem \ref{thm:rec}.  The first is a ``direct'' proof, 
which provides the basis for the sampling algorithm outlined above. 
In addition to the direct proof, we also provide a proof using generating functions
which is seen to be a natural consequence
of MacMahon's development \cite{macmahon15} of symmetric functions,
and yields results of a more general nature.

\subsection{Preliminary observations}
\label{sec:prelim}

For $\r\in\N^n$, let $\r'$ denote the conjugate of $\r$, that is, $r_i'=\#\{j: r_j\geq i\}$ for $i = 1,2,3,\dotsc$. For $\r,\s\in\N^\infty$, let $\r\wedge\s$ denote the component-wise minimum, that is, $(r_1\wedge s_1,r_2\wedge s_2,\dotsc)$. In particular, $\r\wedge\1 =(r_1\wedge 1,r_2\wedge 1,\dotsc)$.
Recall our convention that $\Nn$ is considered to be the subset of
$\N^\infty$ such that all but the first $n$ components are zero. (Similarly, we consider $\Z^n\sbs\Z^\infty$.)

\begin{lem} Let $\u,\v\in\Nn$ such that $\u\leq\v$.
\label{lem:equivalence}
\enum{
\item Suppose $\s\in\N^d$ for some $d\in\N$. Then $\overline{\v-\u} =\bar\v\reduce\s$ if and only if $\s = \v'-(\v-\u)'$.
\item If $\s = \v'-(\v-\u)'$ then $\sum s_i =\sum u_i$ and $\s\leq\bar\v + L\s$.
\item If $\s = \v'-(\v-\u)'$ and $\u\leq\v\wedge\1$ then $\s\leq\bar\v$.
}
\end{lem}
\begin{pf} (1) Letting $I$ be the identity operator, a straightforward calculation shows that for any $d\in\N$, $\r\in\Z^d$, we have $\left(\sum_{k = 0}^\infty L^k\right)(I-L)\r =\r$ and $(I-L)\left(\sum_{k = 0}^\infty L^k\right)\r =\r$, that is, $(I-L)^{-1} =\sum_{k=0}^\infty L^k$ on $\Z^d$, where $I$ is the identity operator. Further, $(I-L)^{-1}\bar\r =\r'$. Thus, $\overline{\v-\u} =\bar\v\reduce\s =\bar\v-\s+L\s$ if and only if $(I-L)\s =\bar\v-\overline{\v-\u}$ if and only if $\s =( I-L)^{-1}(\bar\v-\overline{\v-\u}) =\v'-(\v-\u)'$.

(2) If $\s = \v'-(\v-\u)'$ then $\sum s_i =\sum v_i'-\sum(\v-\u)_i'=\sum v_i-\sum(v_i-u_i) =\sum u_i$ since $\sum r_i'=\sum r_i$ for all $\r\in\N^n$. By (1), $\bar\v-\s+L\s =\bar\v\reduce\s =\overline{\v-\u}\geq\0$ and so $\s\leq\bar\v+L\s$.

(3) If $\s = \v'-(\v-\u)'$ and $\u\leq\v\wedge\1$ then by definition, $s_i =\#\{j: v_j\geq i\}-\#\{j: v_j-u_j\geq i\} =\#\{j: v_j = i\And u_j = 1\}\leq\#\{j: v_j = i\} =\bar v_i$.
\end{pf}

\begin{lem} Let $\v\in\Nn$, $k\in\N$, and let $f(\u)=\v'-(\v-\u)'$ for $\u\in\N^n$ such that $\u\leq\v$.
\label{lem:maps}
\enum{
\item $f(C^{\v\wedge\1}(k)) = C^{\bar\v}(k)$, and for any $\s\in C^{\bar\v}(k)$, $\#\{\u\in C^{\v\wedge\1}(k): f(\u) =\s\} ={\bar\v\choose\s}$.
\item $f(C^\v(k)) =\{\s :\s\in C^{\bar\v + L\s}(k)\}$, and for any $\s$ such that $\s\in C^{\bar\v + L\s}(k)$, $\#\{\u\in C^\v(k): f(\u) =\s\} ={\bar\v + L\s\choose\s}$.
}
\end{lem}
\begin{pf} (1) $f(C^{\v\wedge\1}(k))\sbs C^{\bar\v}(k)$ follows from Lemma \ref{lem:equivalence}(2 and 3). Let $\s\in C^{\bar\v}(k)$. Choose $\u$ as follows. For $i = 1,2,3,\dotsc$, choose $s_i$ of the $\bar v_i$ positions $j$ such that $v_j = i$, and set $u_j = 1$ for each chosen $j$. (Set $u_j = 0$ for all remaining $j$.) 
This determines some $\u\in C^{\v\wedge\1}(k)$ such that $s_i =\#\{j: v_j = i\And u_j = 1\}$ for all $i$. Furthermore, it is not hard to see that any such $\u$ is obtained by such a sequence of choices. Now, as in the proof of Lemma \ref{lem:equivalence}(3), $s_i =\#\{j: v_j = i\And u_j = 1\}$ if and only if $f(\u) =\s$ (when $\u\leq\v\wedge\1$). Hence, $f(C^{\v\wedge\1}(k))\sps C^{\bar\v}(k)$, and since there were ${\bar\v\choose\s}$ possible ways to choose $\u$, then this proves (1).

(2) $f(C^\v(k))\sbs\{\s:\s\in C^{\bar\v+L\s}(k)\}$ follows from Lemma \ref{lem:equivalence}(2). Suppose $\s\in C^{\bar\v+L\s}(k)$. Let $\r =\bar\v$ and $\ttt =\r\reduce\s$. Note that $\ttt\geq\0$ since $\s\leq\r + L\s$. Also, $\r,\s,\ttt\in\N^d$ where $d =\max v_j$. Choose $\u$ as follows. First, consider the $r_d$ positions $j$ in $\v$ at which $v_j = d$. There are ${r_d\choose t_d}$ ways to choose $t_d$ of these $r_d$ positions. Having made such a choice, we set $u_j = v_j-d = 0$ for each such $j$ that was chosen. Next, consider the $r_{d-1}$ positions $j$ at which $v_j = d-1$, in addition to the $r_d-t_d$ remaining positions at which $v_j = d$. There are ${r_{d-1}+(r_d-t_d)\choose t_{d-1}}$ ways to choose $t_{d-1}$ of these. Having made such a choice, we set $u_j = v_j-(d-1)$ for each such $j$ that was chosen. Continuing in this way, for $i = d-2,\dotsc, 1$: consider the $r_i$ positions $j$ in $\v$ which $v_j = i$, in addition to the $r_{i +1} +\cdots + r_d-t_d-\cdots-t_{i +1}$ remaining positions at which $v_j>i$, choose $t_i$ of these (in one of $\ds{r_i + r_{i +1} +\cdots + r_d-t_d-\cdots-t_{i +1}\choose t_i}$ ways), and set $u_j = v_j-i$ for each such $j$ that was chosen. After following these steps for each $i$, set $u_j = v_j$ for any remaining positions $j$. This determines some $\u$ such that $\0\leq\u\leq\v$.

Now, for $i = d,d-1,\dotsc,1$, we have chosen $t_i$ positions $j$ and we have set $u_j = v_j-i$. That is, $t_i =\#\{j: v_j-u_j = i\}$, and so $\ttt =\overline{\v-\u}$. Hence, $\overline{\v-\u} =\bar\v\reduce\s$ (by the definition of $\ttt$), so $\s = f(\u)$ by Lemma \ref{lem:equivalence}(1), and additionally, $\sum u_j =\sum s_j = k$ by \ref{lem:equivalence}(2). Thus, we have shown that $f(C^\v(k))\sps\{\s:\s\in C^{\bar\v+L\s}(k)\}$.

Using $t_j = r_j-s_j+s_{j+1}$ (the definition of $\ttt$), we see that there were
$$
{r_d\choose t_d}{r_{d-1}+(r_d-t_d)\choose t_{d-1}}\cdots{r_1+ r_2+\cdots + r_d-t_d-\cdots-t_2\choose t_1}
$$
$$
={r_d\choose s_d}{r_{d-1}+s_d\choose s_{d-1}}\cdots{r_1+ s_2\choose s_1} ={\r+L\s\choose\s}>0
$$
ways to make such a sequence of choices, where the inequality holds since $\s\leq\r+L\s$. Hence, there are at least ${\r+L\s\choose\s}$ distinct choices of $\u\in C^\v(k)$ such that $f(\u) =\s$. On the other hand, given any $\u\in C^\v(k)$ such that $f(\u) =\s$, we have $\ttt =\overline{\v-\u}$ (by Lemma \ref{lem:equivalence}(1)), thus $t_i =\#\{j: u_j = v_j-i\}$, and since $v_j\geq i$ for any $j$ such that $u_j = v_j-i$, such a $\u$ is obtained by one of the sequences of choices above. Hence, $\#\{\u\in C^\v(k): f(\u) =\s\} ={\bar\v+L\s\choose\s}$.
\end{pf}

\subsection{Direct proof}
\label{sec:direct} 

We are now prepared to prove Theorem \ref{thm:rec}. Recall the statement of the theorem:

{\it
The number of matrices with margins $(\p,\q)\in\Nm\x\Nn$ is given by
\enum{
\item $\ds\bN(\p,\r)=\sum_{\s\in C^\r(p_1)} {\r\choose\s} \bN(L\p,\r\reduce\s)$
\hspace{6mm} for binary matrices, and
\item $\ds\bM(\p,\r)=\sum_{\s\in C^{\r+L\s}(p_1)} {\r+L\s \choose\s} \bM(L\p,\r\reduce\s)$
\hspace{6mm} for $\N$-valued matrices,
}
where $\r=\bq$, and in \textup{(2)}, we sum over all $\s$ such that $\s\in C^{\r+L\s}(p_1)$.
}

{\bf Proof of Theorem 2.1} 

(1) First, we prove the binary case. Let $(\p,\q)\in\Nm\x\Nn$, $\r=\bq$. 
Using Lemma \ref{lem:maps}(1), define the surjection $f: C^{\q\wedge\1}(p_1)\to C^\r(p_1)$ by $f(\u) =\q'-(\q-\u)'$. Then
$$
\bN(\p,\r)=N(\p,\q) \eq{a} \sum_{\u\in C^{\q\wedge\1}(p_1)} N(L\p,\q-\u)$$
$$\eq{b} \sum_{\s\in C^\r(p_1)} \sum_{\u \in f\inv(\s)} N(L\p,\q-\u)
\eq{c} \sum_{\s\in C^\r(p_1)}  {\r\choose\s} \bN(L\p,\r\reduce\s).
$$
Step (a) follows from partitioning the set of $(\p,\q)$ matrices 
according to the first row $\u\in C^{\q\wedge\1}(p_1)$ of the matrix.
Step (b) partitions $C^{\q\wedge\1}(p_1)$ into the level sets of $f$, that is, the sets $f\inv(\s)=\{\u\in C^{\q\wedge\1}(p_1) : f(\u)=\s\}$
as $\s$ ranges over $f(C^{\q\wedge\1}(p_1))=C^\r(p_1)$.
Step (c) follows since if $f(\u)=\s$ then $\overline{\q-\u}=\r\reduce\s$ (by Lemma \ref{lem:equivalence}(1)) and thus
$N(L\p,\q-\u)=\bN(L\p,\r\reduce\s)$, and since $\#f\inv(\s)={\r\choose\s}$ (by Lemma \ref{lem:maps}(1)) .
This proves \ref{thm:rec}(1).

(2) Now, we consider the $\N$-valued case.
Let $S=\{\s : \s\in C^{\r+L\s}(p_1)\}$. 
Using Lemma \ref{lem:maps}(2), define the surjection $g: C^\q(p_1)\to S$ by $g(\u) =\q'-(\q-\u)'$. 
Then, similarly,
$$
\bM(\p,\r)=M(\p,\q) \eq{a} \sum_{\u\in C^\q(p_1)} M(L\p,\q-\u)$$
$$\eq{b} \sum_{\s\in S} \sum_{\u \in g\inv(\s)} M(L\p,\q-\u)
\eq{c} \sum_{\s\in S}  {\r+L\s\choose\s} \bM(L\p,\r\reduce\s).
$$
As before, step (a) follows from partitioning the set of matrices 
according to the first row $\u\in C^\q(p_1)$, 
step (b) partitions $C^\q(p_1)$ into the level sets of $g$,
and step (c) follows since $\#g\inv(\s)={\r+L\s\choose\s}$ (by Lemma \ref{lem:maps}(2)).
This proves Theorem \ref{thm:rec}. \qed

\subsection{Generating function proof}

In addition to the direct approach above, one may also view the recursions
as the application of a certain differential operator to a certain symmetric functions.
Although such operators were used extensively by MacMahon \cite{macmahon15}
on problems of this type, at first it would appear that for computation this approach
would be hopelessly inefficient in all but the simplest examples.
In fact, it turns out that a simple
observation allows one to exploit regularities in the present problem, 
reducing the computation time to polynomial for bounded margins.
Specifically, when there are many columns with the same sum,
the symmetric function under consideration has many repeated factors, and 
the action of the operator in this situation takes a simplified form.

We will identify $N(\p,\q)$ and $M(\p,\q)$ as the coefficients of certain symmetric
functions, introduce an operator for extracting coefficients,
and show that its action yields the recursion above.

Let $e_n$ denote the elementary symmetric function of degree $n$, in a
countably infinite number of variables $\{x_1,x_2,\dotsc\}$:
$$
e_n:=\sum_{r_1<r_2<\cdots<r_n} x_{r_1} x_{r_2} \cdots x_{r_n},
$$
and let $h_n$ be the complete symmetric function of degree $n$:
$$
h_n:=\sum_{r_1\le r_2\le\cdots\le r_n}x_{r_1}x_{r_2}\cdots x_{r_n},
$$
where $r_1,\dotsc,r_n\in\{1,2,3,\dotsc\}$.  
For convenience, let $x_0=e_0=h_0=1$ and $e_n=h_n=0$ if $n<0$.
Given $\r\in\Nn$, let $x^\r:=x_1^{r_1}\cdots x_n^{r_n}$ and $x_\r:=x_{r_1}\cdots x_{r_n}$.
Apply the same notation for $e^\r$ and $e_\r$, as well as $h^\r$ and $h_\r$.
Note that if $\r=\bq$, then $x^\r=x_\q$.

\begin{lem}[MacMahon] For any $\p\in\Nm$, $\q\in\Nn$,
\label{lem:coef}
\enum{
\item $N(\p,\q)$ is the coefficient of $x^\p$ in $e_\q$, and
\item $M(\p,\q)$ is the coefficient of $x^\p$ in $h_\q$.
}
\end{lem}
\begin{pf}
The coefficient of $x^\p$ in $e_\q$ is the number of ways to choose one term from each of
the $n$ factors $e_{q_1},\dotsc,e_{q_n}$, such that the product of these terms is $x^\p$.
Observe the correspondence in which 
the $n$ factors in $e_\q=e_{q_1}\cdots e_{q_n}$
are identified with the $n$ columns in the matrix,
and choosing a term $x_\r$ in a given $e_{q_i}$ corresponds to choosing column $i$
to have ones in rows $r_1,\dotsc,r_{q_i}$ (and zeros elsewhere).
For any choice of terms $x_{\r^1},\dotsc,x_{\r^n}$ from 
$e_{q_1},\dotsc,e_{q_n}$ respectively such that 
$x_{\r^1}\cdots x_{\r^n}=x^\p$, we have a binary matrix with margins $(\p,\q)$, and 
conversely, for any such matrix there is such a choice of terms $x_{\r^1},\dotsc,x_{\r^n}$.
Thus, the coefficient of $x^\p$ in $e_\q$ is also the number of such matrices, $N(\p,\q)$.

The proof for $M(\p,\q)$ is the same, except in this case, 
choosing a term $x^\r$ in a given $h_{q_i}$ corresponds to choosing column $i$
to have entries $r_1,\dotsc,r_m$, and a sequence such that
$x^{\r^1}\cdots x^{\r^n}=x^\p$ corresponds to an $\N$-valued matrix with margins $(\p,\q)$.
\end{pf}

In what follows, when we say ``series'', we mean a formal power series in $x_1,x_2,\dotsc$.
Write $\xx=(x_1,x_2,\dotsc)$ for the sequence of variables, and let $R\xx=(0,x_1,x_2,\dotsc)$.
For $k\in\N$, define the differential operator:
$$
D_k:=\frac{1}{k!}\frac{\del^k}{\del x_1^k}\Big |_{\xx=R\xx}.
$$
In other words, after taking the $k$th derivative with respect to $x_1$
and dividing by $k!$, replace $x_1$ with zero, and $x_{i+1}$ with $x_i$ for $i=1,2,\dotsc$.
Acting on a series in $x_1,x_2,\dotsc$, the operator $D_k$ annihilates every term except
those in which the power of $x_1$ is exactly $k$.
(Note that $D_k$ coincides with Hammond's operator \cite{hammond82,macmahon15}, 
on any symmetric series.)
\ifthenelse {\boolean{details}} {
\begin{detail}
$D_k$ coincides with Hammond's operator on any symmetric series.
\end{detail}
\begin{pf}
Let $d_1:=\frac{d}{de_1}+e_1\frac{d}{de_2}+e_2\frac{d}{de_3}+\cdots$, 
and $H_k:=\frac{1}{k!}(d_1^k)$, where $(d_1^k)$ indicates the algebraic product
$d_1\x d_1\x \cdots \x d_1$ (as opposed to successive application, denoted $(d_1)^k$),
e.g. $$(d_1^2)=\sum_{i,j=1}^{\infty} e_{i-1} e_{j-1} \frac{d^2}{de_i de_j}.$$
The Hammond operator is $H_k$.  We show that $D_k=H_k$ on the symmetric series.

It suffices to show that $D_p e^\r=H_p e^\r$ holds for any $p\in\N$, $\r\in\Nm$.
For, any symmetric series (polynomial, resp.) $f$
is a series (polynomial) in the elementary symmetric series (polynomials),
and since $D_p$ and $H_p$ are linear, then this will show that $D_pf=H_pf$.
If $\s\in C_m(p)$, then
$$\frac{d^p}{de_1^{s_1} \cdots de_m^{s_m}}e_1^{r_1} \cdots e_m^{r_m}
=\prod_{i=1}^m \frac{d^{s_i}}{de_i^{s_i}} e_i^{r_i}
=\prod_{i=1}^m (r_i)_{s_i} e_i^{r_i-s_i},$$
and thus
$$H_p e^\r=
\frac{1}{p!} \sum_{\s\in C_m(p)} {p \choose s_1,\dotsc,s_m} e_0^{s_1}\cdots e_{m-1}^{s_m} 
\frac{d^p}{de_1^{s_1} \cdots de_m^{s_m}}e_\r$$
$$=\sum_{\s\in C_m(p)} {\r \choose \s} \prod_{i=1}^m e_i^{r_i-s_i}e_{i-1}^{s_i}
  =\sum_{\s\in C^\r(p)} {\r \choose \s} e^{\r\reduce\s}=D_p e^\r,$$
the second-to-last step following since ${\r \choose \s}=0$ unless $\s\leq\r$, and
the last step following by Lemma \ref{lem:h2} below.
\end{pf}
} {}
Define 
$$
D_\r:=D_{r_n}\cdots D_{r_1}
$$
(note the reverse order) where $n =\max\{j: r_j\neq 0\}$ if $\r\neq\0$ and $D_\r$ is the identity operator otherwise.
By applying the operator $D_\r$, we keep only terms exactly divisible by $x^\r$
(that is, the power of $x_i$ is $r_i$ for $i=1,\dotsc,n$).
In particular, if $f$ is a homogeneous series of degree $\sum r_i$,
(so that each term has degree $\sum r_i$), then 
$D_\r f$ is a number equal to the coefficient of $x^\r$ in $f$.
Since $e_\q$ and $h_\q$ are homogeneous series of degree $\sum q_i$, then by Lemma \ref{lem:coef} we have
\begin{cor}\
\label{cor:NM} For any $\p\in\Nm$, $\q\in\Nn$ such that $\sum p_i =\sum q_i$,
\enum{
\item $D_\p e_\q=N(\p,\q)$,
\item $D_\p h_\q=M(\p,\q)$. \qed
}
\end{cor}

The following identities begin to reveal the utility of the operators $D_k$.
\begin{lem}[MacMahon] For $n,k\in\N$, 
\label{lem:ham}
\enum{
\item $D_k h_n=h_{n-k}$
\item $D_k e_n=\branch{e_{n-k}}{k\le 1}{0}{k>1}$
\item For any functions $f_1,\dotsc,f_n$,
$$D_k(f_1\cdots f_n)=\sum_{\s\in C_n(k)} (D_{s_1}f_1) \cdots (D_{s_n}f_n).$$
}
\end{lem}
\begin{pf}
(1) and (2) are straightforward calculations. 
For (3), writing $\partial^k =\frac{\partial^k}{\partial x_1^k}$, we have
$$
k! D_k(f_1\cdots f_n) 
=\sum_{\r\in\{1,\dotsc,n\}^k}(\partial^{\bar r_1} f_1)\cdots(\partial^{\bar r_n} f_n)\Big |_{\xx=R\xx}
$$
$$
=\sum_{\s\in C_n(k)}\frac{k!}{s_1!\dotsc s_n!}(\partial^{s_1} f_1)\cdots(\partial^{s_n} f_n)\Big |_{\xx=R\xx}
= \sum_{\s\in C_n(k)} k! (D_{s_1}f_1) \cdots (D_{s_n}f_n),
$$
where the first step follows by recursive
application of the product rule, and the second by collecting like terms.
\end{pf}

\nnl

\begin{lem}[Power rules] For any $k\in\N$, $\r\in\Nn$,
\label{lem:h2}
\enum{
\item $\ds D_k e^\r=\sum_{\s\in C^\r(k)} {\r\choose\s} e^{\r\reduce\s}$
\item $\ds D_k h^\r=\sum_{\s\in C^{\r+L\s}(k)} {\r+L\s \choose\s} h^{\r\reduce\s}.$
}
\end{lem}
\begin{pf} (1) For any $m,i\in\N$, 
$$
D_k e_i^m={m \choose k} e_i^{m-k} e_{i-1}^k
$$
by Lemma \ref{lem:ham}(2 and 3). Thus,
$$D_k e^\r
=D_k (e_1^{r_1}\cdots e_n^{r_n})
\eq{a} \sum_{\s\in C_n(k)} (D_{s_1}e_1^{r_1}) \cdots (D_{s_n}e_n^{r_n})$$
$$\eq{b} \sum_{\s\in C_n(k)} \left({r_1 \choose s_1}e_1^{r_1-s_1}e_0^{s_1}\right)
\cdots \left({r_n \choose s_n}e_n^{r_n-s_n}e_{n-1}^{s_n}\right)$$
$$\eq{c} \sum_{\s\in C_n(k)} {\r\choose\s} e_1^{r_1-s_1+s_2} e_2^{r_2-s_2+s_3}
\cdots e_n^{r_n-s_n}
\eq{d} \sum_{\s\in C^\r(k)} {\r\choose\s} e^{\r\reduce\s},$$
with (a) by Lemma \ref{lem:ham}(3), (b) by the preceding observation, (c) by collecting factors, and (d) since
${\r \choose \s}=0$ unless $\s\leq \r$ and by the definition of $\r\reduce\s$.

(2) Let $m =\sum r_i$ and let $\v\in\N^m$ be any vector such that $\bar\v =\r$, so that $h^\r = h_\v$. Let $S=\{\s : \s\in C^{\r+L\s}(k)\}$, and using Lemma \ref{lem:maps}(2), define the surjection $g: C^\v(k)\to S$ by $g(\u) =\v'-(\v-\u)'$. Then
$$
D_k h^\r=D_k(h_{v_1}\dotsc h_{v_m}) \eq{a} \sum_{\u\in C_m(k)}(D_{u_1} h_{v_1})\dotsc(D_{u_m} h_{v_m})
$$
$$
\eq{b}\sum_{\u\in C_m(k)} h_{\v-\u} \eq{c} \sum_{\u\in C^\v(k)} h_{\v-\u} =\sum_{\u\in C^\v(k)} h^{\overline{\v-\u}}
$$
$$
\eq{d} \sum_{\s\in C^{\r+L\s}(k)}\sum_{\u\in g^{-1}(\s)} h^{\overline{\v-\u}}\eq{e}\sum_{\s\in C^{\r+L\s}(k)}{\r+L\s\choose\s} h^{\r\reduce\s},
$$
where (a) follows from Lemma \ref{lem:ham}(3), (b) by \ref{lem:ham}(1), (c) since $h_j = 0$ if $j<0$ and thus $h_{\v-\u} = 0$ if $\u\not\leq\v$, (d) by \ref{lem:maps}(2), and (e) by \ref{lem:equivalence}(1) and \ref{lem:maps}(2).
\end{pf}

We now complete the generating function proof of Theorem \ref{thm:rec}.
If $\p\in\Nm$, $\q\in\Nn$, $\sum p_i =\sum q_i$, and $\r=\bq$, then by Lemma \ref{lem:h2}(1),
$$D_\p e^\r=D_{L\p}(D_{p_1}e^\r)
=\sum_{\s\in C^\r(p_1)} {\r\choose\s} D_{L\p} e^{\r\reduce\s},$$
and since $e^\r=e_\q$, then using Corollary \ref{cor:NM} (twice) we have
$$\bN(\p,\r)=N(\p,\q)=D_\p e_\q=D_\p e^\r
=\sum_{\s\in C^\r(p_1)} {\r\choose\s} \bN(L\p,\r\reduce\s).$$
This proves Theorem \ref{thm:rec}(1).
Similarly, in view of Corollary \ref{cor:NM}, Theorem \ref{thm:rec}(2) follows immediately from Lemma \ref{lem:h2}(2).

\nnl
\nl

\section{Computation time}
\label{sec:comp}

Let $W(\r):=\sum_{k=1}^n kr_k=$ the {\it weight} of $\r\in\Z^n$.

\begin{lem}[Properties of the weight] If $\r,\s\in\Z^n$ then
\label{lem:W}
\enum{
\item $W(\r+\s)=W(\r)+W(\s)$
\item $W(\s-L\s)=\sum s_i$
\item $W(\r\reduce\s)=W(\r)-\sum s_i$
\item $W(\bar\s) = \sum s_i$.
}
\end{lem}
\begin{pf}
All four are simple calculations.
\end{pf}

For the rest of this section, fix $(\p,\q)\in\Nm\x\Nn$ such that $\sum p_i=\sum q_i$, and
consider $(\p,\q)$ to be the margins of a set of $m\x n$ matrices. 
First, we address the time to compute $N(\p,\q)$ using Algorithm \ref{alg:c}, 
and $M(\p,\q)$ will follow easily.

Let $\D(\p,\q)$ denote the set of nontrivial nodes $(\u,\bv)$ in the directed acyclic graph 
(as discussed in Section \ref{sec:rec})
descending from $(\p,\bq)$ (including $(\p,\bar\q)$), where nontrivial means $(\u,\bv)\neq(\0,\0)$.
Let $\d_k(j):=\{\s\in\N^k : W(\s)=j\}$ for $j,k\in\N$.
The intuitive content of the following lemma is that the graph descending from $(\p,\bq)$ is
contained in a union of sets $\d_k(j)$ with weights decreasing by steps of 
$p_1,\dotsc,p_m$.

\begin{lem}[Descendants] $\D(\p,\q) \sbs \{(\u,\bv) : \u=L^{j-1}\p, \, \bv\in\d_b(t_j),\,j=1,\dotsc,m\}$,
where $t_j =\sum_{i=j}^m  p_i$ and $b =\max q_i$.
\label{lem:L}
\end{lem}
\begin{pf}
By the form of the recursion, $(\u,\bv)\in\D(\p,\q)$ if and only if for some $1\le j\le m$ there exist
$\s^1,\dotsc,\s^{j-1}$ in $C^{\r^1}(p_1),\dotsc,C^{\r^{j-1}}(p_{j-1})$ respectively, with
$\r^1=\bq$, $\r^{i+1}=\r^i\reduce\s^i$ for $i=1,\dotsc,j-1$, such that
$(\u,\bv)=(L^{j-1}\p,\r^j)$.
For $j\geq 2$, by Lemma \ref{lem:W}(3 and 4),
$$W(\r^j)=W(\r^{j-1}\reduce\s^{j-1})=W(\r^{j-1})-p_{j-1}=W(\r^{j-2})-p_{j-2}-p_{j-1}$$
$$=\cdots=W(\r^1)-(p_1+\cdots+p_{j-1})=\sum_{i = 1}^n q_i-\sum_{i = 1}^{j-1} p_i = 
\sum_{i = 1}^m p_i-\sum_{i = 1}^{j-1} p_i = t_j,$$
and $\r^j\in\N^b$ by construction, so
$\r^j\in\d_b(t_j)$. 
Hence, $(\u,\bv)=(L^{j-1}\p,\r^j)$ belongs to the set as claimed.
\end{pf}

Let $T(\p,\q)$ be the time (number of machine operations) required by the algorithm 
(Algorithm \ref{alg:c}) to compute $N(\p,\q)$ after precomputing all needed binomial coefficients.
Let $\t(\u,\bv)$ be the time to compute $\bN(\u,\bv)$ given
$\bN(L\u,\bv\reduce\s)$ for all $\s\in C^\bv(u_1)$.  That is, $T(\p,\q)$ is the
time to perform the entire recursive computation, whereas $\t(\u,\bv)$ is the time to perform
a given call to the algorithm not including time spent in subcalls to the algorithm.

Let $n_0 :=\#\{i: q_i>0\}$ denote the number of nonempty columns.
By constructing Pascal's triangle, we precompute all possible binomial coefficients that will be needed, and store them in a lookup table. We only need binomial coefficients with entries less or equal to $n_0$, for the following reason. In the binary case, the recursion involves numbers of the form ${\bar\v\choose\s}$ with $\s\leq\bar\v$, and for any descendent $(\u,\bar\v)$ and any $i>0$ we have $\bar v_i\leq n_0$ since the number of columns with sum $i$ is less or equal to the total number of nonempty columns. For the $\N$-valued case, the same set of binomial coefficients will be sufficient, since then we have numbers of the form ${\bar\v + L\s\choose\s}$ with $\s\leq\bar\v + L\s$, and thus
$$
\bar v_i + s_{i+1}\leq\bar v_i + \bar v_{i+1} + s_{i+2}\leq \cdots \leq\bar v_i + \bar v_{i+1} +\bar v_{i+2} + \cdots \leq n_0,
$$
where the last inequality holds because the number of columns $j$ with sum greater or equal to $i$ is no more than the total number of nonempty columns. Since the addition of two $d$-digit numbers takes $\Theta(d)$ time, and there are ${n_0 +2\choose 2}$ binomial coefficients with entries less or equal to $n_0$, then the bound $\log {j \choose k} +1 \leq n_0\log 2 +1 $ on the number of digits for such a binomial coefficient shows that this pre-computation can be done in $\O(n_0^3)$ time. Except in trivial cases (when the largest column sum is 1), the additional time needed does not affect the bounds on $T(\p,\q)$ that we will prove below.

We now bound the time required for a given call to the algorithm.

\begin{lem}[Time per call] $\t(\u,\bv)\leq\O((ab+c)(\log c)^3 |C_b(u_1)|)$ for $(\u,\bv)\in\D(\p,\q)$,
where $a =\max p_i, b =\max q_i$, and $c =\sum p_i$.
\label{lem:t}
\end{lem}
\begin{pf}
Note that we always have $\b v_i\leq n_0$, since the number
of columns with sum $i$ cannot exceed the number of nonempty columns.
Thus, in the recursion formula,
for each $\s$ in the sum corresponding to $(\u,\bv)$, we have the bound
$${\bv \choose \s} = \prod_{i=1}^b {\b v_i\choose s_i} 
\leq \prod_{i=1}^b \b v_i^{s_i} \leq n_0^{\sum_i s_i} \leq n_0^a\leq c^a.$$ 
Let $T_m(k)$ be the time required to multiply two numbers of magnitude $k$ or less.
By the Sch\"{o}nhage-Strassen algorithm \cite{schonhage71}, 
$T_m(k)\leq \O((\log k)(\log\log k)(\log\log\log k))$.
Therefore, $T_m({\bv \choose \s})\leq T_m(c^a)\leq\O(a(\log c)^3)$.
Since we have precomputed the binomial coefficients, the time required 
to compute ${\bv \choose \s}$ is thus bounded by $\O(ab(\log c)^3)$.
To finish computing the term corresponding to $\s$ in the recursion formula, we must multiply
 ${\bv \choose \s}$ by $\bN(L\u,\bv \reduce \s)$.
Since 
$$\bN(L\u,\bv \reduce \s)\leq N(\p,\q)\leq \prod_{i=1}^m {n_0\choose p_i}\leq \prod_{i=1}^m n_0^{p_i} = n_0^c\leq c^c,$$
then this multiplication can be done in
$T_m(N(\p,\q))\leq T_m(c^c) \leq \O(c(\log c)^3)$ time.
Since we are summing over $C^\bv(u_1)$, and $C^\bv(u_1)\sbs C_b(u_1)$, then altogether we have
$\t(\u,\bv) \leq \O((ab+c)(\log c)^3 |C_b(u_1)|)$ for the time per call.
\end{pf}

\begin{lem}[Bound on weighted simplices] $\ds\#\d_k(j)\leq {j+k-1 \choose k-1}$ for any $j,k\in\N$.
\label{lem:D}
\end{lem}
\begin{pf}
The map $f(\r)=(1r_1,2r_2,\dotsc,kr_k)$ is an injection $f:\d_k(j)\to C_k(j)$.
Thus, $\#\d_k(j)\leq \#C_k(j)={j+k-1 \choose k-1}$.
\end{pf}

\nl
We are now ready to prove Theorem \ref{thm:eff} for the case of $N(\p,\q)$.

\nl
{\bf Proof of Theorem \ref{thm:eff} for $N(\p,\q)$}

By storing intermediate results in a lookup table,
once we have computed $\bN(\u,\bv)$ upon our first visit to
node $(\u,\bv)$, we can simply reuse the result for later visits.
Hence, we need only expend $\t(\u,\bv)$ time for each node $(\u,\bv)$ occuring
in the graph.  Let $t_j =\sum_{i=j}^m  p_i$ and $d=(ab+c)(\log c)^3$. Then
$$T(\p,\q)=\sum_{(\u,\bv)\in\D(\p,\q)} \t(\u,\bv)
\leql{a} \sum_{j=1}^m \sum_{\bv\in\d_b(t_j)} \t(L^{j-1}\p,\bv)$$
$$\leql{b} \sum_j \sum_\bv \O(d |C_b(p_j)|)
 = \sum_j \O(d|C_b(p_j)| |\d_b(t_j)|)$$
$$\leql{c} \sum_j \O(d {p_j+b-1 \choose b-1}{t_j+b-1 \choose b-1})$$
$$\leql{d} \sum_j \O(d {a+b-1 \choose b-1}{c+b-1 \choose b-1})
\leq \O(dm (a+b-1)^{b-1}(c+b-1)^{b-1}),$$
where (a) follows by Lemma \ref{lem:L}, (b) by \ref{lem:t},
(c) by \ref{lem:D}, and
(d) since $p_j\leq a$ and $t_j\leq c$.
This proves (1) and (2).  
Now, (3) and (4) follow from (2) since $a \leq c \leq bn$.
\qed

\nnl
{\bf Proof of Theorem \ref{thm:eff} for $M(\p,\q)$}

Other than the coefficients, the only difference between the recursion for $\bM(\p,\bq)$ and 
that for $\bN(\p,\bq)$ is that we are summing over $\s$ such that $\s\in C^{\r+L\s}(p_1)$.
Lemma \ref{lem:L} holds with the same proof, except with
$C^{\r^1}(p_1),\dotsc,C^{\r^{j-1}}(p_{j-1})$
replaced by $C^{\r^1+L\s^1}(p_1)$, $\dotsc$, $C^{\r^{j-1}+L\s^{j-1}}(p_{j-1})$, respectively.
Considering Lemma \ref{lem:t},
let $(\u,\bv)$ be a descendent of $(\p,\bq)$ in the graph for $\bM(\p,\bq)$, and
let $\s$ be such that $\s\in C^{\bv+L\s}(u_1)$. 
Similarly to before, recalling that $\bar v_i + s_{i+1}\leq n_0$ (as proven above in our discussion of precomputing the binomial coefficients), we have
$${\bv+L\s \choose \s} = \prod_{i=1}^b {\b v_i+s_{i+1} \choose s_i} 
\leq \prod_i (\b v_i+s_{i+1})^{s_i} \leq n_0^{\sum s_i} \leq n_0^a\leq c^a.$$ 
This yields $T_m({\bv+L\s \choose \s})\leq T_m(c^a)\leq\O(a(\log c)^3)$, just as before. Since 
$$\bM(L\u,\bv \reduce \s)\leq M(\p,\q)\leq \prod_{i=1}^m {p_i+n_0-1\choose p_i}\leq \prod_i (2c)^{p_i} =(2c)^c,$$
then we also obtain $T_m(M(\p,\q))\leq T_m((2c)^c) \leq \O(c(\log c)^3)$ as before.
Further, $\{\s:\s\in C^{\bv+L\s}(u_1)\}\sbs C_b(u_1)$, so altogether the time per call is
$\O((ab+c)(\log c)^3 |C_b(u_1)|)$, and thus the result of Lemma \ref{lem:t} continues to hold.
With this result, the proof of the bounds goes through as well.
\qed

This completes the proof of Theorem \ref{thm:eff}.
Now, we address the time required to uniformly sample a matrix with specified margins.
Let $T_r(k)$ be the maximum over $1\leq j\leq k$ of the expected time to generate a random integer uniformly between $1$ and $j$.
If we are given a random bitstream (independent and identically distributed Bernoulli$(1/2)$ random variables) with constant cost per bit, then $T_r(k) =\O(\log k)$, since for any $j \leq k$, $\lceil \log_2 j \rceil \leq \lceil \log_2 k \rceil$ random bits can be used to generate an integer uniformly between $1$ and $2^{\lceil \log_2 j\rceil}$ and then rejection sampling can be used to generate uniform samples over $\{1,\dotsc,j\}$. Since the expected value of a Geometric$(p)$ random variable is $1/p$, then the expected number of samples required to obtain one that falls in $\{1,\dotsc,j\}$ is always less than $2$. 
More generally, for any fixed $d\in\N$, if we can draw uniform samples from $\{1,\dotsc,d\}$, then we have $T_r(k) =\O(\log k)$ by considering the base-$d$ analogue of the preceding argument. 

\begin{lem}[Sampling time] 
\label{lem:stime}
Algorithm \ref{alg:s} takes $\O(mT_r(n^c)+maT_r(n)+mb\log(a+b))$ expected time per sample in the binary case, and
$\O(mT_r((2c)^c)+maT_r(n)+mb\log(a+b))$ expected time per sample in the $\N$-valued case.
If $T_r(k) = \O(\log k)$, then this is $\O(mc\log c)$ expected time per sample in both cases.
\end{lem}
\begin{rk} If $b$ is bounded then $\O(mc\log c)\leq\O(mn\log n)$ since $c\leq bn$, and so this is polynomial expected time for bounded column sums.
\end{rk}
\begin{pf}
By the form of the recursion, the depth of the graph descending from $(\p,\bq)$
is equal to the number of rows $m$, since $\p\in\Nm$ and thus $L^m \p=\0$.
For each of the $m$ iterations of the sampling algorithm, we begin at some node $(\u,\bv)$,
and we must (A) randomly choose a child $(L\u,\bv\reduce\s)$ with probability proportional to its 
count times the number of corresponding rows, and then
(B) choose a row uniformly from among the ${\bv\choose\s}$ possible choices in the binary case
(or ${\bv+L\s\choose\s}$ in the $\N$-valued case).

First consider the binary case.
To randomly choose a child, consider a partition of the integers $1,\ldots,N(\u,\v)$
with each part corresponding to a term in the recursion formula for $\bN(\u,\bv)$.
Generate an integer uniformly at random between $1$ and $N(\u,\v)$, and choose the corresponding child.
Generating such a random number takes $T_r(N(\u,\v))\leq T_r(N(\p,\q))\leq T_r(n_0^c)$ time.
Since there are no more than ${a+b-1 \choose b-1}\leq (a+b-1)^{b-1}$ children at any step,
one can determine which child corresponds to the chosen number in $\O((b-1)\log(a+b-1))$ time
by organizing the children in a binary tree.
So (A) takes $\O(T_r(n_0^c)+b\log(a+b))$ time.
Choosing a row consists of uniformly sampling a subset of size $s_i$ from a set of $\b v_i$ elements,
for $i=1,\dotsc,b$.
Sampling such a subset can be done by sampling without replacement $s_i$ times,
which takes $\sum_{j=0}^{s_i-1} T_r(\b v_i -j) \leq s_i T_r(n_0)$ time.
So (B) can be done in $\sum_{i=1}^b s_i T_r(n_0) \leq a T_r(n_0)$ time.
Repeating this process $m$ times, once for each row, we see that sampling a matrix takes
$\O(mT_r(n_0^c)+mb\log(a+b)+ma T_r(n_0))$ time. 
If $T_r(k) \leq \O(\log k)$, this is 
$\O(mc \log n_0 + mb\log(a+b)+ma \log n_0) \leq \O(m c\log c)$ since $a,b,n_0\leq c$.

For the $\N$-valued case, the same argument applies, 
replacing $N(\u,\v)$ with $M(\u,\v)$, $n_0^c$ with $(2c)^c$, and $\b v_i$ with $\b v_i+s_{i+1}$.
\end{pf}

\bibliography{refs}
\bibliographystyle{amsplain}

\appendix
\section{Enumeration results}
{\bf Binary matrices with margins   
$(70,30,20,10,5^{(6)},4^{(10)},3^{(20)},2^{(60)}),(4^{(80)},3^{(20)})$}

860585058801817078819959949756041558231879514104670757612387\\
280341919502865086909993523205599348663646837362726765460951\\
032776118129432733489342067673016169716787054236343091407458\\
802261593735765113169808512677339861494709092492858489355535\\
514748397544147637928475318462070009855280569561693514768239\\
201499080842592443823774161366680107327323365049702068246736\\
456919918589686056321467354298509024976141650428747522863473\\
529515269318246400000000000000000000000

\nnl
{\bf $\N$-valued matrices with margins   
$(70,30,20,10,5^{(6)},4^{(10)},3^{(20)},2^{(60)}),(4^{(80)},3^{(20)})$}

620017488391049592297896956531192562528805388295441812965295\\
130897484012791595142882674755488640101825726867156331426482\\
441148514978852842582445295040041143220637964258279947442682\\
896809706562683189375098411751981435132377208717294759756041\\
358372207736032818841045369779439398975681041714752821787419\\
816573563436066161167632677774184809010338787868042742993719\\
703936093873250600121874335524794990013547042810153560084573\\
133035731217642637607153615611029851392000000000000000000000\\
000

\nnl
{\bf Ehrhart polynomials $H_n(r)$ for $n=4,\dotsc,8$}
 
$H_4(1) = 24$ \\
$H_4(2) = 282$ \\
$H_4(3) = 2008$ 

$H_5(1) = 120$ \\
$H_5(2) = 6210$ \\
$H_5(3) = 153040$ \\
$H_5(4) = 2224955$ \\
$H_5(5) = 22069251$ \\
$H_5(6) = 164176640$ 

$H_6(1) = 720$ \\
$H_6(2) = 202410$ \\
$H_6(3) = 20933840$ \\
$H_6(4) = 1047649905$ \\
$H_6(5) = 30767936616$ \\
$H_6(6) = 602351808741$ \\
$H_6(7) = 8575979362560$ \\
$H_6(8) = 94459713879600$ \\
$H_6(9) = 842286559093240$ \\
$H_6(10) = 6292583664553881$ 

$H_7(1) = 5040$ \\
$H_7(2) = 9135630$ \\
$H_7(3) = 4662857360$ \\
$H_7(4) = 936670590450$ \\
$H_7(5) = 94161778046406$ \\
$H_7(6) = 5562418293759978$ \\
$H_7(7) = 215717608046511873$ \\
$H_7(8) = 5945968652327831925$ \\
$H_7(9) = 123538613356253145400$ \\
$H_7(10) = 2023270039486328373811$ \\
$H_7(11) = 27046306550096288483238$ \\
$H_7(12) = 303378141987182515342992$ \\
$H_7(13) = 2920054336492521720572276$ \\
$H_7(14) = 24563127009195223721952590$ \\
$H_7(15) = 183343273080700916973016745$ 

$H_8(1) = 40320$ \\
$H_8(2) = 545007960$ \\
$H_8(3) = 1579060246400$ \\
$H_8(4) = 1455918295922650$ \\
$H_8(5) = 569304690994400256$ \\
$H_8(6) = 114601242382721619224$ \\
$H_8(7) = 13590707419428422843904$ \\
$H_8(8) = 1046591482728407939338275$ \\
$H_8(9) = 56272722406349235035916800$ \\
$H_8(10) = 2233160342369825596702148720$ \\
$H_8(11) = 68316292103293669997188919040$ \\
$H_8(12) = 1667932098862773837734823042196$ \\
$H_8(13) = 33427469280977307618866364694400$ \\
$H_8(14) = 562798805673342016752366344185200$ \\
$H_8(15) = 8115208977465404874100226492575360$ \\
$H_8(16) = 101857066150530294146428615917957029$ \\
$H_8(17) = 1128282526405022554049557329097252992$ \\
$H_8(18) = 11161302946841260178530673680176000200$ \\
$H_8(19) = 99613494890126594335550124219924540800$ \\
$H_8(20) = 809256770610540675454657517194018680846$ \\
$H_8(21) = 6031107989875562751266116901999327710720$

\end{document}